\documentclass[12pt]{amsart}

\usepackage{graphicx}
\usepackage{subfigure}
\usepackage{amssymb,xy}
 \xyoption{all}

\topmargin by -\headsep \textheight 8.9in \oddsidemargin 0pt
\evensidemargin \oddsidemargin \marginparwidth 0.5in \textwidth
6.5in

\begin{document}
\begin{center}
{\large {\bf Recombining binomial tree for  constant elasticity of variance process}}\\
${ }$\\
Hi Jun Choe, Jeong Ho Chu and So Jeong Shin\\
Department of Mathematics, Yonsei University, Seoul, Republic of Korea\\
\end{center}
${ }$\\

{\small{\bf ABSTRACT.}}  ${ }$
The theme in this paper is the recombining binomial tree to price
American put option when the underlying stock follows constant
elasticity of variance(CEV) process. Recombining nodes of binomial
tree are decided from finite difference scheme to emulate CEV
process and the tree has a linear complexity. Also it is derived
from the differential equation the asymptotic envelope of the
boundary of tree. Conducting  numerical experiments, we confirm the
convergence and accuracy of the pricing by our recombining binomial
tree method.
As a result, we can compute the price of American put option under CEV model, effectively.\\
${ }$



\section{\bf INTRODUCTION}

 Black and Scholes\cite{BLACK} derived the celebrated option pricing
formula under the assumption that underlying stock price follows
Geometric Brownian Motion(GBM). Under this assumption the
distribution of prices is lognormal and the volatility is constant.
However, the empirical evidence does not support the assumptions in
the lognormal distribution and the constant volatility. In other
words, unlike the basic assumption of GBM, we observe that the
implied volatility embedded in the market price of option changes
according to the exercise price and expiration. This phenomenon is
called `volatility smile'. CEV model can explain the `volatility
smile' phenomenon and can more closely approximate the real world
than GBM. So CEV model has been a popular alternative stock process
model and there have been many attempts for financial applications.
The option pricing formula when
the underlying stock process follows CEV model was derived by Cox
and Ross\cite{COX}. The financial implications of the CEV model are studied by
Beckers\cite{BECKERS}. It was noted that CEV model can fit the
volatility skew of stock options. Indeed, Cox and Ross\cite{COX} and
Emanuel and MacBeth\cite{EMANUEL} derived a closed-form solution for
European option when a positive elasticity is assumed.
Schroder\cite{SCHRODER} simplified the formula. However computations
involving the non-central chi-square distribution function are
complicated, and as an attempt Schroder introduced an analytic
approximation of option pricing for CEV model. The explicit formula
is only useful for European vanilla options, not for American
options and other exotic options. Therefore it is common practice
that American option is computed by binomial tree method. Nelson and
Ramaswamy\cite{NELSON} suggested a simple binomial process
approximation to describe CEV process. But, as noted by Nelson and
Ramaswamy (\cite{NELSON}, p.418), their simple binomial process
approximation proposed deteriorates as maturity is lengthened. To
overcome such a computational burden, we propose a real simple and
accurate binomial tree to estimate the value of American put options
and so on. The novelty of our binomial tree is exact recombination
for CEV model. From finite difference scheme for partial
differential equation a recombining tree is made for CEV model.

Moreover it is well known that the early exercise valuation problem
can be solved by the binomial tree method. The binomial tree is an
efficient and powerful method for pricing American options in
contrast to the partial differential equation method and other
numerical methods such as Monte-Carlo simulation. Tomer
Neu-Ner\cite{TomerNeuNer} discussed alternative methods of pricing
and compared them with the binomial method. He claimed that the
binomial tree method is an extremely valuable tool for option
pricing under CEV model. The remainder of this paper is organized as
follows. In section 2, we review briefly CEV model. Section 3 is the
main part in this paper. First, we derive a partial differential
equation which holds for any type of option. Second, we build a
binomial tree to approximate the CEV process and to evaluate the
American put option valuation. In other words, we introduce the
structure of binomial tree which can exactly recombine. In section
4, we present numerical results and discuss about the convergence of
binomial process built in chapter 3. In section 5, we compute
American put option value under the CEV model by the recombining
binomial tree. In the final section, we present the conclusion of
this paper.

\section{\bf CONSTANT ELASTICITY OF VARIANCE MODEL}

CEV model was proposed by Cox and Ross\cite{COX} as an alternative
to the Black and Scholes\cite{BLACK} model(GBM). This model
proposes the following relationship between stock price $ S $ and
volatility $\upsilon(S,t)$
$$  \upsilon(S,t) = \sigma S^{{\beta -2 } \over 2}. $$
It means that the elasticity of return variance with respect to
stock price $ S $ equals $ \beta -2 $.
$$ {{d \upsilon^2 / \upsilon^2 } \over {dS/S}}=  \beta - 2 . $$
In CEV model, the stock price $ S $ is assumed to be governed by the
diffusion process:
$$ dS = \mu S dt + \sigma S^{\beta \over 2} dW. $$
Here, we denote the stock price at an instant of time $t$ as $ S$,
the change in the stock price over the increment $dt$ as $dS$.
$\mu$, $\sigma$ and $\beta$ are positive constants. $  dW $  is
Wiener process. We assume the stock pays no dividends.

 If $ \beta =2 $, then the volatility $ \sigma(S,t) $ is $ \sigma $.
So in this case, CEV model is just GBM model. Otherwise, observe
that volatility varies with moves in the stock price level and time.
If $ \beta >2 $, the volatility and stock price move in the same
direction. If $ \beta <2 $, the volatility increases as the stock
price decreases. In this case, the probability distribution is
similar to that observed for stock option with a heavy left tail. It
is known, based on empirical data, that stock prices and volatility
have an inversely relationship. So we only consider the situation
when $ 0 < \beta < 2 $.

\section{\bf BINOMIAL TREE FOR CEV DIFFUSION }

\subsection{FINITE DIFFERENCE METHOD FOR BLACK-SCHOLES EQUATION}

We consider the general stock process.
$$ dS = b(S,t)dt + \sigma(S,t)dW. $$
First, define a function  $ V(S,t) $ that gives the option value for
an asset price $S\ge0$  at any time $ t $ with $0\le t \le T$. The
key idea is hedging to eliminate risk. We can obtain the following
equation by doing similar arguments to obtain a Black-Scholes
equation:
\begin{equation}\label{BLACK}
{\partial V \over \partial t} + {\partial V \over
\partial S}{r S} + {1 \over 2} { \partial^2 V \over \partial S^2}
{\sigma^2(S,t) } - {rV} = 0,
\end{equation}
where $r$ is the risk free rate and it is constant and positive. We
are to recognize that time goes backward in (\ref{BLACK}).

Second, we apply FDM to the above equation. FDM is a straightforward
method for solving Partial Differential Equations(PDE). FDM requires
the domain to be replaced by a grid. The key step in deriving FDM is
to replace differential operators with finite difference operators.
By plugging the difference formula into the PDE (\ref{BLACK}), a
difference equation (\ref{FDE}) is obtained:
\begin{equation}\label{FDE}
{{{V^n_i}-{V^{n-1}_i}} \over \Delta t} +
{{{V^n_{i+1}}-{V^n_{i-1}}} \over {{S^n_{i+1}}-{S^n_{i-1}}}}{r S^n_i}
+ {{1 \over 2}{\sigma^2(S^n_i)}} {{{{V^n_{i+1} - V^n_i} \over
{S^n_{i+1} - S^n_i}}-{{V^n_i - V^n_{i-1}} \over { S^n_i -
S^n_{i-1}}}} \over {{1 \over 2} {(S^n_{i+1}  - S^n_{i-1})}}} - r
V^{n-1}_i =0.
\end{equation}

 Here, $V^n_i$ denotes the value of the option corresponding to asset price $S^n_i$ at $(n,i)$ node. The superscript indicates the
time level.
\begin{figure}
    \begin{center}
    $$\xymatrix{  & V_{i-1}^n \ar@{-}[l] \ar@{-}[r] \ar@{-}[dr] & V_i^n \ar@{-}[d] \ar@{-}[r] & V_{i+1}^n \ar@{-}[r] \ar@{-}[dl]& \\ &V_{i-1}^{n-1} \ar@{-}[l] \ar@{-}[r]&V_{i}^{n-1} \ar@{-}[r] & V_{i+1}^{n-1} \ar@{-}[r] & }$$
    \end{center}
    \caption{ Finite Difference Method}
\end{figure}
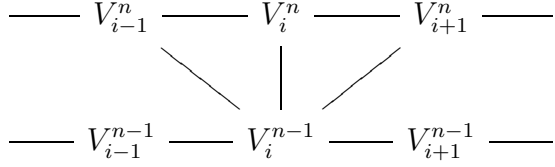

By simplifying we obtain
$$(1+r \Delta t) V^{n-1}_i = V^n_i  + {{V^n_{i+1} -
V^n_{i-1}} \over {S^n_{i+1} - S^n_{i-1}}} { r \Delta t S^n_i} +
{{{{\sigma^2 (S^n_i) \Delta t} } \over {S^n_{i+1} - S^n_{i-1}}}
\left({{V^n_{i+1} - V^n_i} \over {S^n_{i+1} -S^n_i}} - {{V^n_i
-V^n_{i-1}} \over {S^n_i - S^n_{i-1}}}\right)}$$
\begin{eqnarray*}
 &=& \left({{r \Delta t S^n_i} \over {S^n_{i+1} - S^n_{i-1}} } + {
{\sigma^2(S^n_i)\Delta t} \over {(S^n_{i+1} -
S^n_{i-1})(S^n_{i+1} - S^n_i )} } \right)V^n_{i+1} \\
&&+ \left( 1 - { {\sigma^2(S^n_i) \Delta t} \over ( S^n_{i+1}
-S^n_{i-1})}
 ( {1 \over {{S^n_{i+1} -S^n_i }} }+ {{1 \over {S^n_i  - S^n_{i-1}}}})\right)V^n_i \\
&&  + \left({{-r \Delta t S^n_i} \over {S^n_{i+1} - S^n_{i-1}} }
  + { {\sigma^2(S^n_i)\Delta t} \over {(S^n_{i+1} - S^n_{i-1})(S^n_i - S^n_{i-1} )} }
  \right)V^n_{i-1}.
  \end{eqnarray*}
  So, we have the explicit form of $ V^{n-1}_i $  as following
 \begin{equation}\label{ITER}
  V^{n-1}_i = {1 \over { 1 + r \Delta t}} [ h^n_{i+1} V^n_{i+1} + h^n_i V^n_i + h^n_{i-1}
  V^n_{i-1}], \end{equation}
where \begin{equation*}
 h^n_{i+1} = {{ r \Delta t
S^n_i} \over {S^n_{i+1} - S^n_{i-1}} }
  + { {\sigma^2(S^n_i)\Delta t} \over {(S^n_{i+1} - S^n_{i-1})(S^n_{i+1} - S^n_i )} }
  \end{equation*}
\begin{equation*} h^n_i =  1 - { {\sigma^2(S^n_i) \Delta t} \over ( S^n_{i+1}
-S^n_{i-1})}
 ( {1 \over {{S^n_{i+1} -S^n_i }} }+ {{1 \over {S^n_i  -
 S^n_{i-1}}}}) \end{equation*}
\begin{equation*} h^n_{i-1} = {{-r \Delta t S^n_i} \over {S^n_{i+1} - S^n_{i-1}}
}
  + { {\sigma^2(S^n_i)\Delta t} \over {(S^n_{i+1} - S^n_{i-1})(S^n_i - S^n_{i-1} )}
  }.
  \end{equation*}
If the finite difference scheme corresponds to the binomial tree, we
have to make $ h^n_i = 0 $, that is,
\begin{equation}\label{DT}  1 - {
{\sigma^2(S^n_i) \Delta t} \over ( S^n_{i+1} -S^n_{i-1})}
 ( {1 \over {{S^n_{i+1} -S^n_i }} }+ {{1 \over {S^n_i  -
 S^n_{i-1}}}}) = 0 . \end{equation}
We observe that, if $   h^n_i =0 $  then  $ h^n_{i+1} + h^n_{i-1} = 1 $.\\

 \subsection{STRUCTURE OF THE BINOMIAL TREE}

 In CEV model $ \sigma(S,t)= \sigma S^{\beta \over 2} $ and by
simplifying the equation (\ref{DT}), we obtain the essential
recombination equation
\begin{equation}\label{rrr} (S^n_{i+1} - S^n_i)(S^n_i
-S^n_{i-1})=\sigma^2 {S^n_i}^\beta \Delta t.
\end{equation}

Now, we build a recombining binomial tree of stock prices. The basic
idea of binomial tree construction is as follows. Here, we let
$S(i,j)$(=$S^i_j$) denote the price of stock at i-time level
($j=1,2,\cdots,2i-1$). Put $S(i,j)=S(i-1,j-1)$, $i=2,\cdots,n$, $ j
= 2,\cdots, 2i-2 $. And $ S(i,1)$ and
 $S(i, 2i-1)$ are determined from the equation (\ref{rrr}).
 \begin{figure}
 \begin{center}
 $$ \xymatrix{&&&S(1,1) \ar[dl] \ar@{.>}[dd] \ar[dr]&&& \\
&&S(2,1) \ar[dl] \ar@{.>}[dd] \ar[dr] && S(2,3) \ar[dl] \ar@{.>}[dd] \ar[dr] && \\
&S(3,1) \ar[dl] \ar@{.>}[d] \ar[dr] &&S(3,3)\ar[dl] \ar@{.>}[d] \ar[dr] && S(3,5)\ar[dl] \ar@{.>}[d] \ar[dr] &\\
S(4,1) &&S_(4,3) && S(4,5) && S(4,7) }
$$
 \end{center}
 \caption{Structure of binomial tree : exactly recombining}
  \end{figure}
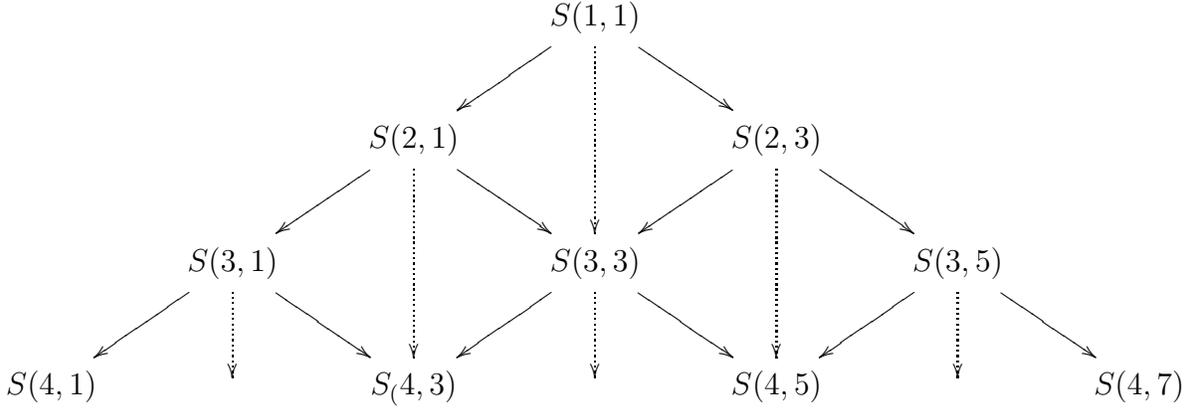

Note that $S(i,j)$ is underlying stock price at $i$-th time step. We
explain the procedure in detail. First, $S(1,1)$ is the current
stock price. Put $ S(2,2)=S(1,1)$. If  $S(2,3)$ is given, then two
known values $S(2,3)$ and $S(1,1)(=S(2,2))$, and one unknown value
$S(2,1)$ should satisfy the recombination equation (\ref{rrr})
because the stock price follows CEV model. That is, the value
$S(2,1)$ is determined by the two known values $S(1,1)(=S(2,2))$,
$S(2,3)$ and the equation (\ref{rrr}).

Now, put $ S(3,3)=S(2,2)$ and $S(3,4)=S(2,3)$. Again, the value
$S(3,5)$ is determined by the two known values $S(3,3)$,
$S(2,3)(=S(3,4))$ and the equation (\ref{rrr}). Similarly we obtain
$S(3,1)$ as determined by $S(3,3)$, $S(2,1)(=S(3,2))$ and the
equation (\ref{rrr}).

We describe one more step. Put  $S(4,5)=S(3,4)$, $ S(4,6)=S(3,5)$.
We obtain $ S(4,7) $ as determined by the equation (\ref{rrr})
 with the two known values $S(4,5)$ and $S(3,5)(=S(4,6))$ plugged
 in.
Similarly, putting $S(4,3)=S(3,2)$ and $S(4,2)=S(3,1)$, we obtain
$S(4,1)$ by plugging the two known values $S(4,3)$ and
$S(3,1)(=S(4,2))$ into the equation (\ref{rrr}).
Observe that $S(4,3)$, $S(3,3)(=S(4,4))$, $S(4,5)$ satisfy the
equation (\ref{rrr}).
 Continuing in the same manner, we can build a binomial tree of stock
 price of CEV model. Observe that, once $S(2,3)$ has been determined, the binomial tree is uniquely determined.

The largest benefit of the binomial tree constructed in this manner
is as follows. This is a most natural and simplest binomial tree
that allows exact recombining under CEV model. Cox \& Rubinstein(\cite{COXRUB}, p362) have constructed a binomial approximation for
the CEV diffusion. However, it turns out that computation is not
appropriate in their case because tree does not recombine and thus
the number of nodes doubles at each time step. When the binomial
tree does not recombine at each node, the computation is not
efficient. On the other hand, the binomial tree in which recombining
occurs at each level is efficient and speedy to compute because the
number of nodes grows at most linearly with the number of time
intervals. That is, in a recombining binomial process, the stock
price can take $i+1$ possible values after $i$ periods, for
$i=1,2,3,\cdots,n $.

\subsection{FINDING THE FIRST VALUE OF TREE(Determine the increasing rate of the stock price
$u$)}
 Now we have a problem. How can we set the value of $S(2,3)$? In other words, we have to tune the parameter($u$: move up factor) and explain why.
$$dS=\mu S dt + \sigma S^{\beta \over 2} dW .$$
By using the Euler's discretization
\begin{eqnarray*}
  \Delta S&=&\mu S \delta t + \sigma S^{\beta \over 2} Y \sqrt{\Delta t} , \quad\mbox{where}\quad Y \sim N(0,1) \\
   S_{n+1} &=& S_n + \mu S_n \Delta t + \sigma S_n^{\beta \over 2} Y \sqrt{\Delta t}\\
 &=& S_n(1+ \mu \Delta t + \sigma S_n^{{\beta \over 2}-1}Y \sqrt{\Delta
 t}).
 \end{eqnarray*}
Take $Y=1$ so that
$$ S_{n+1} = S_n (1+ \mu \Delta t + \sigma S_n^{{\beta \over 2}-1} \sqrt{\Delta t}).$$
Since $\sqrt{\Delta t}$ is far larger than $\Delta t$ for a small
$\Delta t$, we can ignore the $\Delta t$ term to get
\begin{eqnarray*}
S_{n+1} &\approx& S_n(1+ \sigma S_n^{{\beta \over 2}-1} \sqrt{\Delta t}) \\
 &\approx& S_n e^{\sigma S_n^{{\beta \over 2} -1} \sqrt{\Delta t}}. \end{eqnarray*}
Note that if $\beta=2$,
$$S_{n+1} \approx S_n (1+ \sigma \sqrt{\Delta t})\approx S_n e^{\sigma \sqrt{\Delta t}} $$
So we set the value of $S(2,3)=S(1,1)e^{\sigma S(1,1)^{{\beta \over 2} -1} \sqrt{\Delta t}}$.

\subsection{PROBABILITY OF UPWARD MOVE AT EACH NODE}

 In the CEV model, the volatility is not constant but varies
with the value of the underlying price. When the volatility varies
with the value of the price, the probability of an upward move has
to be recomputed at each node. Now, we compute the probability of an
upward move at each node. For notational simplicity, let  $ \Delta
S^n_i = S^n_{i+1} - S^n_i $. Then the equation (\ref{DT}) becomes
$$ 1 -{{ \sigma^2(S^n_i) \Delta t } \over { \Delta S^n_i + \Delta
S^n_{i-1}} } \left( {1 \over \Delta S^n_i } + { 1 \over \Delta
S^n_{i-1}}\right) = 0. $$
On the other hand, in the process of
binomial tree, we have
$$ V^{n-1}_i  = e^{-r  \Delta t}  ( p^{n-1}_i  V^n_{i+1} +
(1-p^{n-1}_i)V^n_{i-1}). $$
Here, $ p^{n-1}_i $ is the increasing
probability of the stock price at $(n-1, i)$ node. So, we obtain the
following equation by comparing with (\ref{ITER})
\begin{align*}
e^{-r \Delta t} p^{n-1}_i &= {1 \over { 1 +r  \Delta t}} h^n_{i+1}
={1 \over {1 + r \Delta t}} \left({{ r \Delta t S^n_i} \over {\Delta
S^n_i + \Delta S^n_{i-1}} }
  + { {\sigma^2(S^n_i)\Delta t} \over {(\Delta S^n_i +
  \Delta S^n_{i-1})\Delta S^n_i} } \right) \\
e^{-r \Delta t}(1- p^{n-1}_i) &= {1 \over { 1 +r  \Delta t}}
h^n_{i-1} ={1 \over {1 + r \Delta t}} \left({{ - r \Delta t S^n_i}
\over {\Delta S^n_i + \Delta S^n_{i-1}} }
  + { {\sigma^2(S^n_i)\Delta t} \over {(\Delta S^n_i +
  \Delta S^n_{i-1})\Delta S^n_{i-1}} }\right). \end{align*}

Since $ h^n_i =0 $, we also have
\begin{align*}
1 &- { {\sigma^2(S^n_i) \Delta t} \over ( S^n_{i+1} -S^n_{i-1})}
 \left( {1 \over {{S^n_{i+1} -S^n_i }} }+ {{1 \over {S^n_i  -
 S^n_{i-1}}}}\right) = 0 \\
 1 &= {{ \sigma^2(S^n_i) \Delta t } \over { \Delta S^n_i + \Delta
S^n_{i-1}} } \left( {1 \over \Delta S^n_i } + { 1 \over \Delta
S^n_{i-1}}\right) \\
1&= {{{\sigma(S^n_i)}    {\sigma(S^n_i) } } \over { \Delta S^n_i
\Delta S^n_{i-1}}} \Delta t \\
\Delta t&= { \Delta S^n_i \over \sigma(S^n_i)}   { \Delta S^n_{i-1}
\over \sigma(S^n_i)}.
 \end{align*}

 Let $$ { \Delta S^n_i \over \sigma(S^n_i)} = \xi_i ,$$ then with very small
 error, we can write $$ \xi_{i-1} \approx \xi_i =   \chi = \sqrt{\Delta t}.
 $$
 Then the above equation becomes
\begin{eqnarray*} p^{n-1}_i &=&{ {  e^{r \Delta t} } \over {1 + r \Delta t}}
\left({{ r \Delta t S^n_i} \over {\Delta S^n_i + \Delta S^n_{i-1}} }
  + { {\sigma^2(S^n_i)\Delta t} \over {(\Delta S^n_i +
  \Delta S^n_{i-1})\Delta S^n_i} } \right)  \\
 &=&  {{  e^{r \Delta t} } \over {1 + r \Delta t}} {\left(  { 1 \over
2 } r {\sqrt \Delta t } { S^n_i \over \sigma(S^n_i)} +  {1 \over 2 }
\right)}\end{eqnarray*}
\begin{eqnarray*}1- p^{n-1}_i &=&{ { e^{r
\Delta t} } \over {1 + r \Delta t}} \left({{ - r \Delta t S^n_i}
\over {\Delta S^n_i + \Delta S^n_{i-1}} }
  + { {\sigma^2(S^n_i)\Delta t} \over {(\Delta S^n_i +
  \Delta S^n_{i-1})\Delta S^n_{i-1}} }\right) \\
  &=&  {{  e^{r \Delta t} } \over {1 + r \Delta t}}
{\left(  - { 1 \over 2 } r {\sqrt \Delta t } { S^n_i \over
\sigma(S^n_i)} +  {1 \over 2 }   \right)}.\end{eqnarray*}

Therefore, in CEV model, we have
\begin{equation}p^n_i =  {{  e^{r \Delta t}
} \over {1 + r \Delta t}} {\left(  { 1 \over 2 } r {\sqrt \Delta t }
{ {S^{n+1}_i}^{1- {\beta \over 2}} \over \sigma} + {1 \over 2 }
\right)}.\end{equation}

\section{\bf NUMERICAL TEST}

In section 3, we built a binomial tree to emulate CEV diffusion. It
provides an efficiency tool to price options because it recombines
exactly. At this point, we present the convergence of pricing by
binomial tree by numerical experiments.

\subsection{EUROPEAN PUT OPTION}

 We have stochastic differential equation representing CEV process as follows:
$$dS=(r-q)Sdt+\sigma S^\alpha dW, $$
where $r$, $q$, and $\alpha$ are parameters of risk free rate,
dividend yield and elasticity, respectively. Under CEV model, the
closed-form formulas for pricing of European call and put options
are available. Cox\cite{COX1975} obtained that
$$c=S_0 e^{-qT} [1- \chi^2(a, b+2, c)]-K e^{-rT} \chi^2 (c,b,a),$$
$$p=K e^{-rT} [1-\chi^2 (c,b,a)]-S_0 e^{-qT}  \chi^2(a, b+2, c),$$
when $\alpha < 1$(or $\beta<2$), and Emanuel and
MacBeth\cite{EMANUEL} derived that
$$c=S_0 e^{-qT} [1- \chi^2(c,-b,a)]-K e^{-rT} \chi^2 (a,2-b,c),$$
$$p=K e^{-rT} [1-\chi^2 (a,2-b,c)]-S_0 e^{-qT}  \chi^2(c,-b,a),$$
when $\alpha > 1$(or $\beta>2$) with \\
$\alpha = {{[Ke^{-(r-q)T}]^{2(1-\alpha)}} \over {(1-\alpha)^2
\omega}}$, $b={1 \over {1-\alpha}}$, $c={{S^{2(1-\alpha)}}\over
{(1-\alpha)^2 \omega}},$ where $\omega = {{\delta^2 \over
{2(r-q)(\alpha-1)}}[e^{2(r-q)(\alpha-1)T}-1] }$ and $\chi^2(z,k,w)$
is the cumulative distribution function of a noncentral chi-square
random variable with noncentrality parameter $\omega$ and $k$
degrees of freedom.

 Now we compare European put option value
between analytic solution and binomial tree solution to check the
convergence of the binomial tree solution.
\begin{table}
   \begin{center}
\begin{tabular}{cccccccccccc}
\hline \hline
\multicolumn{3}{c}{} & \multicolumn{3}{c}{Analytic solution} & \multicolumn{3}{c} {Tree-time step=365}  & \multicolumn{3}{c} {Tree-time stpe 365*2}  \\
\hline
 & $S$ & $E$ & $T$=1/4 & $T$=1/2 & $T$=1 & $T$=1/4 & $T$=1/2 & $T$=1 & $T$=1/4 & $T$=1/2 & $T$=1 \\
\hline
$ \beta$ =0.5 & 0.5 & 1
&0.4876   &0.4753   &0.4447     &0.4859     &0.4719     &0.4450     &0.4858     &0.4719     &0.4449  \\
� &1 &1
& 0.0337    &0.0442     &0.0537     &0.0332     &0.0432     &0.0539     &0.0332     &0.0432     &0.0539  \\
� & 1.5 &1
 &0.0000    &0.0000     &0.0002     &0.0000     &0.0000     &0.0002     &0.0000     &0.0000     &0.0002\\
\hline
$ \beta$=1 & 0.5 &1
&0.4876     &0.4753     &0.4417     &0.4851     &0.4704     &0.4419     &0.4858     &0.4704     &0.4418\\
� & 1 &1
 & 0.0337   &0.0442     &0.0519     &0.0326     &0.0421     &0.0520     &0.0332     &0.0422     &0.0520  \\
� & 1.5 &1
& 0.0000    &0.0000     &0.0003     &0.0000     &0.0000     &0.0003     &0.0000     &0.0000     &0.0003 \\
\hline
$ \beta$ =2 & 0.5 &1
 & 0.4876   &0.4753     &0.4412     &0.4851     &0.4703     &0.4412     &0.4851     &0.4703     &0.4412 \\
� &1 &1
& 0.0337    &0.0442     &0.0487     &0.0316     &0.0403     &0.0487     &0.0316     &0.0403     &0.0488 \\
� & 1.5 &1
& 0.0000    &0.0001     &0.0007     &0.0000     &0.0000     &0.0007     &0.0000     &0.0000     &0.0007 \\
\hline  \hline
\end{tabular}
\end{center}
   \caption{Convergence of European Options under CEV process}
\end{table}
Suppose that $S$, $E$ and $T$ denote current stock price, strike
price and time to maturity, respectively. We use analytic solution
and binomial tree solution to value European put with $S=0.5,1,1.5$,
$E=1$, $ T = {1 \over 4} $,$ 1 \over 2$, $1$, $r=0.05$ and
$\sigma=0.2$. \textit{Analytic Solution} and \textit{Tree} represent
the option values obtained by using the analytic closed form formula
and by using the binomial tree method constructed by this paper,
respectively.  Table 1 shows the result for $n=365$, $n=365 \times
2$ and closed form solution. Observe that with all choice of $n$ the
binomial tree method approximation  \textit{Tree} is close to
\textit{Analytic Solution} at least two decimal places.
 It implies the convergence of the binomial tree built in this paper.
 So we  claim confidently that recombining binomial tree method
 constructed in this paper is good approximation for the solution.

In a different aspect, Nelson and Ramaswamy\cite{NELSON} proposed a
 binomial process approximation for option pricing under CEV model
by using transformation. But as noted by Nelson and
Ramaswamy(\cite{NELSON}, p.418), their binomial process
approximation  deteriorates as maturity is lengthened. Our
recombining binomial tree
 approximates the value of option with linear complexity
although maturity is lengthened. It's simple and efficient.

   We compare the European put option values between closed-form solution and binomial tree method
   by picture. Stock $S$ varies 0 to 3, strike price $E=1$, $T=1$, $r=0.05$ and $\sigma=0.2$.
Also, we know that European put options value is increasing as
$\beta$ is decreasing. Indeed we show the fact by presenting Figure
3 and 4. In Figure 3 and 4, we computed European put options value
as stock price varies $0.5$ to $1.5$.

\begin{figure}
    \begin{center}
    \subfigure[beta=0.5]{\includegraphics[width=0.5\textwidth]       {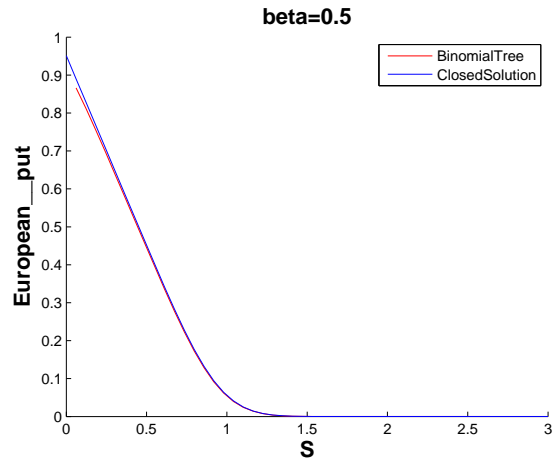}}\\
    \subfigure[beta=1]{\includegraphics[width=0.5\textwidth]{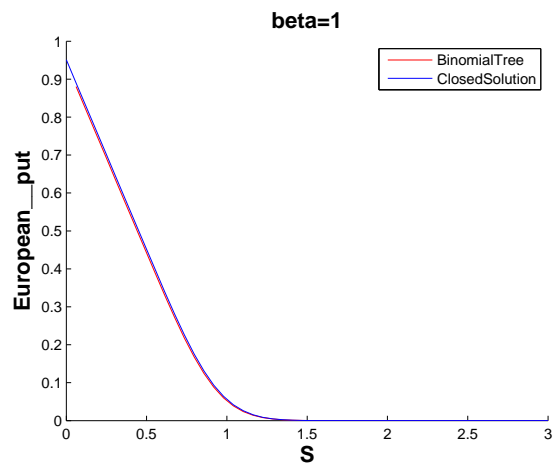}}\\
    \subfigure[beta=2]{\includegraphics[width=0.5\textwidth]{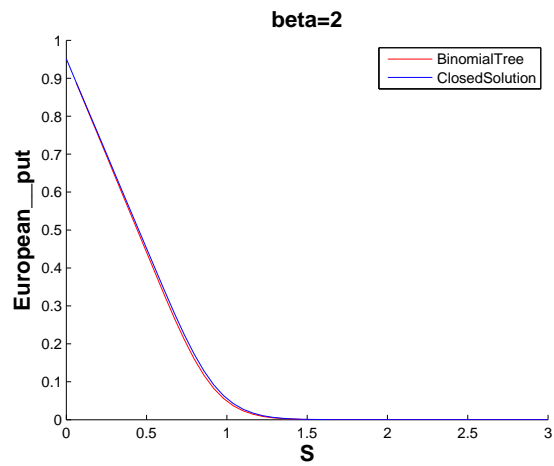}}
    \end{center}
    \caption{ European put option price under the CEV as stock varies}
  \end{figure}

\begin{figure}
  \begin{center}
   \includegraphics[width=1\textwidth]{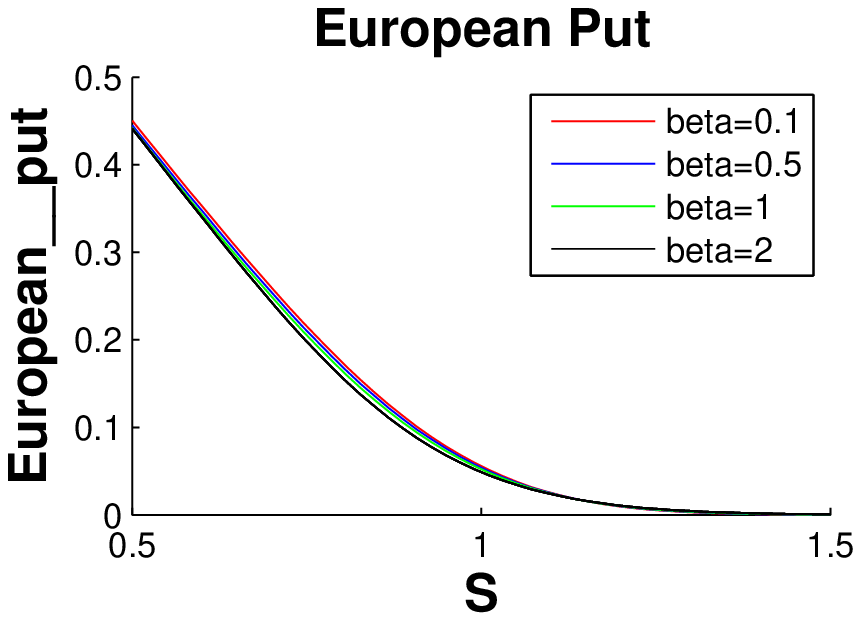}
     \caption{European put option price is increasing as beta decreasing-the value computed via the binomial tree method under the CEV as stock varies}
  \end{center}
Stock $S$ varies $0.5$ to $1.5$, strike price $E=1$, $T=1$, $r=0.05$ and $\sigma=0.2$\\
-red : $\beta=0.1$\\
-blue : $\beta=0.5$\\
-green : $\beta=1$\\
-black : $\beta=2$
\end{figure}

\subsection{ENVELOPE}

In this section we study the range of node set of tree. We find an
asymptotic envelope of boundary of tree.

Let $f^n=S(n,2n-1)$ which is the uppermost branch of tree. Then the
recombination equation (\ref{rrr}) becomes
$$ (f^n - f^{n-1})(f^{n-1}-f^{n-2})=\sigma^2 (f^n)^\beta \Delta t $$
Let different time scale $ \tau ={ t \over{\sqrt{\Delta t}}} $ and
final time $T=N \Delta t $, then $t=n \Delta t$.
$$ f^n \approx y(n \sqrt{\Delta t} ) = y(\tau) $$
Consequently, letting $\Delta t$ go to zero, we obtain the envelope
equation
\begin{align}\label{ENV}
(y'(\tau))^2& = \sigma^2 y(\tau)^\beta  \\
 y(0) &= f^0 .\nonumber
\end{align}
We find easily the solution of the envelope equation (\ref{ENV}):
\begin{align*}
y(\tau) &=exp(\pm \sigma \tau + c) , \quad c=ln(S(1,1)),\quad  \mbox{if}\quad  \beta=2,\\
 y(\tau)&= {{{2-\beta}\over 2}(\pm \sigma \tau + c)}^{2 \over {2-
\beta}}, \quad  c ={{2 \over {2-\beta}}{S(1,1)}^{{2-\beta}\over 2}},
\quad \mbox{if}\quad  0 < \beta < 2 .
\end{align*}



\begin{figure}
    \begin{center}
    \subfigure[beta=1]{\includegraphics[scale=0.7]{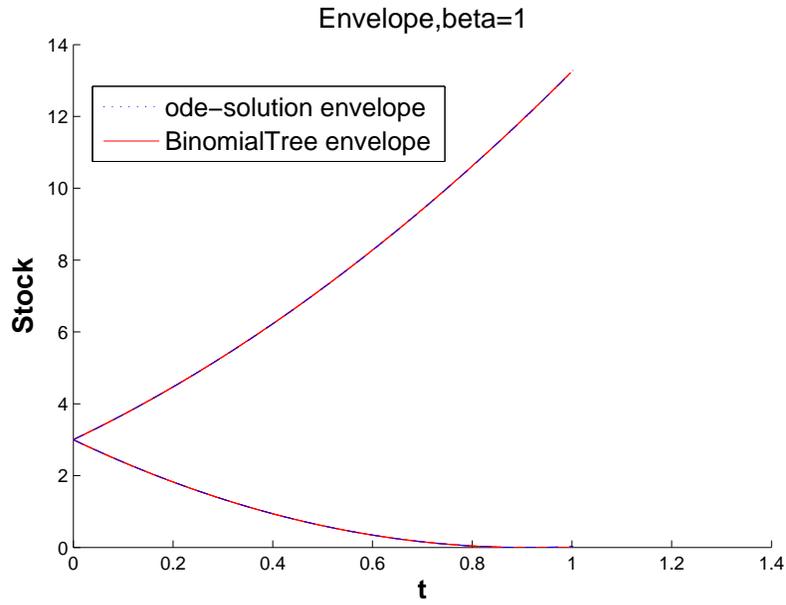}}
    \subfigure[beta=2]{\includegraphics[scale=0.7]{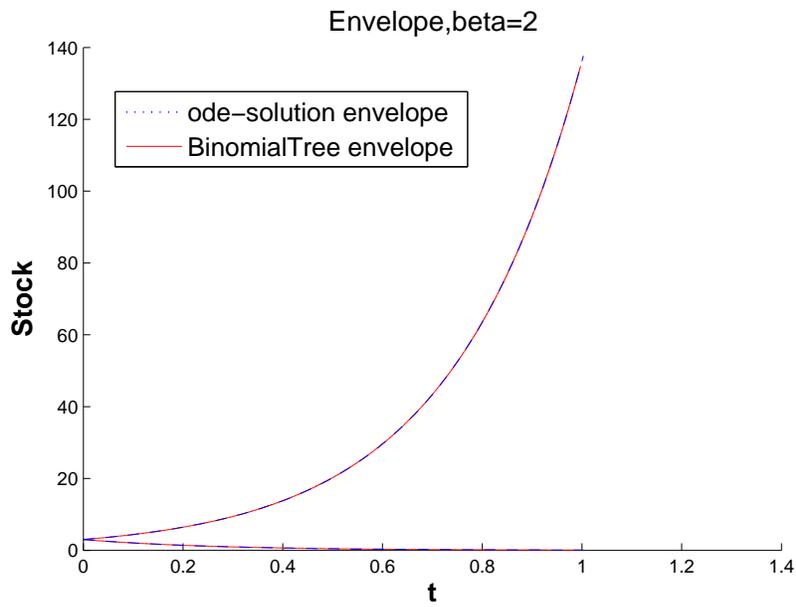}}
    \end{center}
    \caption{ Compare the envelope}
    \begin{flushleft}
     -blue : ode-solution envelope\\
     -red : binomial tree envelope under the situation  and $S=3$, $E=1$, $r=0.05$, $\sigma=0.2$.\\
           (a)$\beta=1$, (b)$\beta=2$
    \end{flushleft}
\end{figure}

We compare the envelope of binomial tree with the analytic envelope.
Figure 5 gives a plot of the envelope when $\beta=1$, $\beta=2$ and
$S=3$, $E=1$, $r=0.05$, $\sigma=0.2$.  The envelopes of binomial
trees follow the asymptotic solutions. There are two graph in each
figure. One(-red) is binomial tree envelope, the other(-blue) is the
solution to the envelope equation (\ref{ENV}). We see that two
curves agree well.

\section{\bf PRICING AMERICAN PUT OPTION UNDER CEV MODEL}

American put option gives its holder the right (but not the
obligation) to sell to the writer a prescribed asset for a
prescribed price at any time between the start date and a prescribed
expiry date in the future. American option differs from European
option by the early exercise possibility. American option can be
exercised at any time between the start date and the expiry date
unlike European option which can only be exercised at maturity.
Unfortunately, there is no analytic solution to the American option
problem in general. It turns out that the binomial tree method can
be used to value American put option. At each node we calculate the
value of the option as a function of the next period prices.
 In chapter $3$, the asset prices in the binomial model under
the CEV diffusion are determined. If the put option is held until
its maturity date $T$, then
\begin{equation} V^N_i = \Lambda(S^N_i).\end{equation}
 Here, $t_N = T$, $  \Lambda(S^n_i)= Max (E - S^n_i,
0) $ and $E$ is an exercise price.
 We work backward through the tree. If the option is retained,
 then $V^n_i$ is $ e^{-r \delta t}(p^n_i V^{n+1}_{i+1} + (1-p^n_i)V^{n+1}_{i-1})$. However,
exercising the option would produce $ \Lambda(S^n_i)$. Hence
choosing the best of the two possibilities leads to the relation.
\begin{equation} V^n_i = Max [ \Lambda (S^n_i),  e^{-r \delta t}(p^n_i V^{n+1}_{i+1} + (1-p^n_i)V^{n+1}_{i-1})].\end{equation}
Then we compute the time zero option value.
\begin{figure}
      \centering
     \includegraphics[width=1\textwidth]{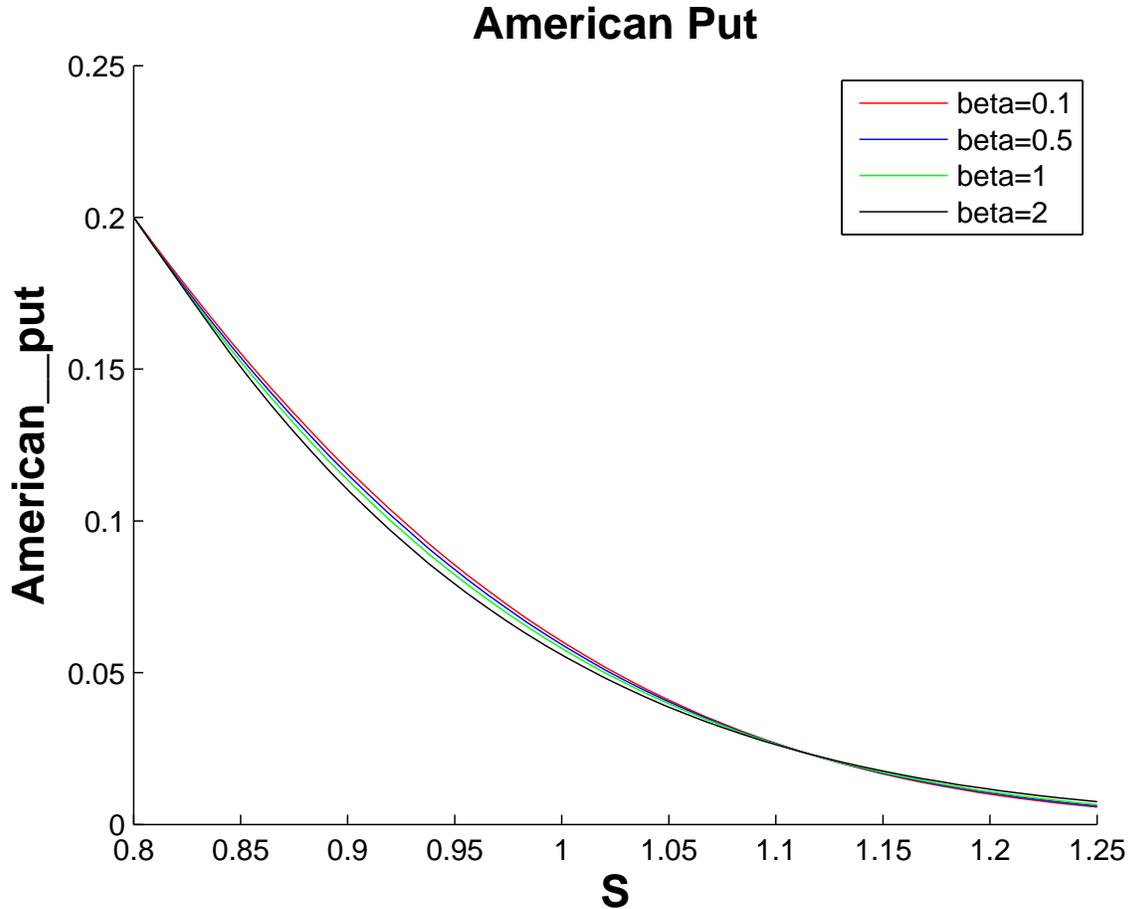}
     \caption{American put option price computed via the binomial tree method under the CEV as stock varies}
     -$E=1$, $T=1$, $r=0.05$ and $\sigma=0.2$ as stock price $S$ varies $0.8$ to $1.25$\\
-red : $\beta=0.1$\\
-blue : $\beta=0.5$\\
-green : $\beta=1$\\
-black : $\beta=2$
\end{figure}
Note $S$, $E$, $T$ denotes current stock price, strike price, time
to maturity, respectively.

In Figure 6, we present numerical value for an American put,
computed by the recombining binomial tree method with $E=1$, $T=1$,
$r=0.05$ and $\sigma=0.2$ as stock price $S$ varies $0.8$ to $1.25$.
Observe that American put option value is increasing as $\beta$ is
decreasing like European put option value.

In figure 7,  we also compare the probability distribution function
of the recombining tree solution and the probability distribution of
analytic solution
$$f(x)= {1 \over {x \sigma \sqrt{2 \pi t}}} exp [{-{log(x/S_0)-(\mu - \sigma^2/2)t}^2 \over {2 \sigma^2 t} }]$$
when $\beta=2$.
\begin{figure}
      \centering
     \includegraphics[width=1\textwidth]{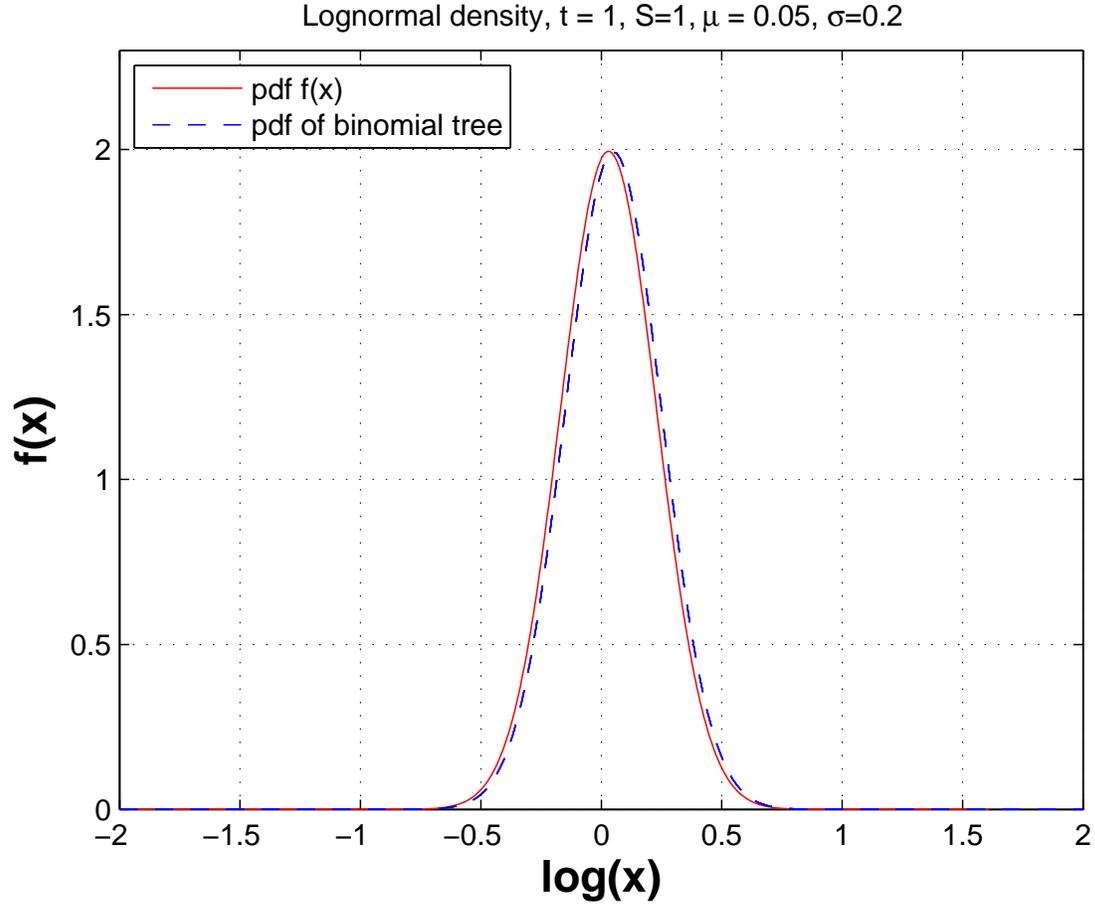}
     \caption{Compare the probability density function of binomial tree ($\beta =2$) to lognormal probability density function}
\end{figure}

\section{\bf CONCLUSION}
In this paper, we discuss the pricing of American put option when
the underlying stock follows Constant Elasticity of Variance (CEV)
process. We constructed a recombining binomial tree to emulate CEV
process. So we can apply it to pricing American put option,
effectively.  We tried to show the convergence of binomial tree
method by comparing the European put option value between analytic
solution and binomial tree. Our numerical results shows a good
convergence. The binomial tree constructed in this paper has the
advantage of being a most natural and simplest because it exactly
recombines and has linear complexity for CEV model. It is our guess
that our idea can be applied to different stochastic processes if we
can derive Black-Scholes type partial differential equation.

\end{document}